\documentclass[12pt,a4paper]{article}
\usepackage{axodraw}
\usepackage{graphicx}
\usepackage{amsmath}
\usepackage{amsfonts}
\usepackage{amssymb}
\newcommand{\vs}{\vspace{-0.25cm}}

\begin{document}
\begin{center}

{\Large
\textbf{Quasi-particle interaction in nuclear matter\\ from chiral
pion-nucleon dynamics}}
\bigskip

{\large N. Kaiser}\\

\bigskip

{\small Physik-Department T39, Technische Universit\"{a}t M\"{u}nchen,
D-85747 Garching, Germany\\
\smallskip

{\it email: nkaiser@ph.tum.de}}

\end{center}

\medskip

\begin{abstract}
Based on a recent chiral approach to nuclear matter we calculate the
in-medium interaction of nucleons at the Fermi surface $|\vec p_{1,2}|=k_f$. 
The isotropic part of this quasi-particle interaction is characterized by four
density-dependent (dimensionful) Fermi-liquid parameters: $f_0(k_f),\,f_0'(k_f
),\, g_0(k_f)$ and $g_0'(k_f)$. In the approximation to $1\pi$-exchange and 
iterated $1\pi$-exchange (which as such leads already to a good nuclear matter 
equation of state) we find a spin-isospin interaction strength of $g_0'(
2m_\pi) = 1.14\, $fm$^2$, compatible with existing empirical values. The 
consistency relations to the nuclear matter compressibility $K$ and the 
spin/isospin asymmetry energies serve as a check on our perturbative 
calculation. In the next step we include systematically the contributions from
$2\pi$-exchange with virtual $\Delta(1232)$-isobar excitation which have been 
found important for good single-particle properties and spin-stability of 
nuclear matter. Without any additional short-distance terms (contributing 
proportional to $k_f^2$) the spin-dependent Fermi-liquid parameters $g_0(
k_{f0})$ and $g'_0(k_{f0})$ come out far too large. Estimates of these 
short-distance parameters from realistic NN-potentials go in the right 
direction, but sizeable enhancement factors are still needed to reproduce 
the empirical values of $g_0(k_{f0})$ and $g_0'(k_{f0})$. This points towards 
the importance of higher order iterations subsumed in the induced interaction. 
We consider also the tensor part of the quasi-nucleon interaction at the Fermi
surface. In comparison to the leading $1\pi$-exchange tensor interaction we
find from the $2\pi$-exchange corrections almost a doubling of the isoscalar
tensor strength $h_0(k_f)$, whereas the isovector tensor strength $h_0'(k_f)$
is much less affected. These features are not changed by the inclusion of the 
chiral $\pi N\Delta$-dynamics. The $l=1$ tensor Fermi-liquid parameters 
$h_1(k_f)$ and $h_1'(k_f)$ follow a similar pattern. 
\end{abstract}

\medskip



\section{Introduction and preparation}
In recent years a novel approach to the nuclear matter problem based on
effective field theory (in particular chiral perturbation theory) has emerged. 
Its key element is a separation of long- and short-distance dynamics and an 
ordering scheme in powers of small momenta. At nuclear matter saturation 
density $\rho_0=2k_{f0}^3/3\pi^2 \simeq 0.16\,$fm$^{-3}$ the Fermi momentum 
$k_{f0}$ and the pion mass $m_\pi$ are comparable scales ($k_{f0}\simeq 2
m_\pi$), and therefore pions must be included as explicit degrees of freedom
in the description of the nuclear many-body dynamics. The contributions to the
energy per particle $\bar E(k_f)$ of isospin-symmetric (spin-saturated)
nuclear matter as they originate from chiral pion-nucleon dynamics have been
computed up to three-loop order in refs.\cite{lutz,nucmat}. Both calculations
are able to reproduce correctly the empirical saturation point of nuclear
matter by adjusting one single parameter (either a contact-coupling $g_0+g_1 
\simeq 3.23$ \cite{lutz} or a cut-off $\Lambda \simeq 0.65\,$GeV \cite{nucmat})
related to unresolved short-distance dynamics.\footnote{The cut-off scale
$\Lambda$ serves in ref.\cite{nucmat} the purpose to tune the strength of an
emerging attractive zero-range NN-contact interaction.} The novel mechanism for
saturation in these approaches is a repulsive contribution to the energy per
particle $\bar E(k_f)$ generated by Pauli blocking in second order (iterated)
one-pion exchange. As outlined in section 2.5 of ref.\cite{nucmat} this
mechanism becomes particularly transparent by taking the chiral limit $m_\pi =
0$. In that case the interaction contributions to  $\bar E(k_f)$ are completely
summarized by an attractive $k_f^3$-term and a repulsive $k_f^4$-term where
the parameterfree prediction for the coefficient of the latter is very close
to the one extracted from a realistic nuclear matter equation of state.  

In a recent work \cite{deltamat} we have extended this chiral approach to 
nuclear matter by including systematically the effects from two-pion exchange 
with single and double virtual $\Delta(1232)$-isobar excitation. The physical
motivation for such an extension is threefold. First, the spin-isospin-3/2 
$\Delta(1232)$-resonance is the most prominent feature of low-energy 
$\pi N$-scattering. Secondly, it is well known that the two-pion exchange 
between nucleons with excitation of virtual $\Delta$-isobars generates the 
needed isoscalar central NN-attraction \cite{gerst} which in phenomenological 
one-boson exchange models is often simulated by a fictitious  scalar 
''$\sigma$''-meson exchange. Thirdly, the delta-nucleon mass splitting 
$\Delta = 293\,$MeV is of the same size as the Fermi momentum $k_{f0} \simeq 
2m_\pi$ at nuclear matter saturation density and therefore pions and 
$\Delta$-isobars should both be treated as explicit degrees of freedom. A
large variety of nuclear matter properties has been investigated in this
extended framework in ref.\cite{deltamat} and it has been found that the 
inclusion of the $\pi N \Delta$-dynamics is able to remove most of the 
shortcomings of previous chiral calculations \cite{nucmat,liquidgas} of
nuclear matter. In particular, the momentum-dependence of the (real)
single-particle potential $U(p,k_{f0})$ near the Fermi surface $p=k_{f0}$ 
improves significantly. The effective nucleon mass $M^* = 0.88 M$ and thus the
density of states at the Fermi surface are now better described. As a
consequence the critical temperature of the liquid-gas phase transition is
lowered to the realistic value $T_c=15\,$MeV. Moreover, the isospin properties
improve also substantially by the inclusion of the chiral $\pi
N\Delta$-dynamics, as exemplified by the density dependence of the neutron
matter equation of state $\bar E_n(k_n)$ and the asymmetry energy $A(k_f)$
(see Figs.\,10,11 in ref.\cite{deltamat}).

Given the fact that both groundstate and single-particle properties can be 
well described by perturbative chiral $\pi N$-dynamics up to three-loop order,
it is natural to consider in a further step the in-medium interaction of 
quasi-nucleons (at the Fermi surface). Following this program we will
therefore calculate in this work the density dependent Fermi-liquid parameters
as they emerge from perturbative chiral pion-nucleon dynamics. 

Fermi-liquid theory was invented by Landau \cite{landau} to describe strongly 
interacting (normal) many-fermion systems at low temperatures. At low 
excitation energies the elementary excitations of a many-fermion system are 
long-lived quasi-particles which in a certain sense interact weakly. The 
quasi-particles can thought of as free particles dressed by the 
interactions with the dense many-body medium. Landau's Fermi-liquid theory has 
been applied successfully to liquid $^3$He \cite{babu}, nuclear matter and 
nuclei \cite{migdal}. In the latter case a set of Fermi-liquid parameters, 
describing the particle-hole interaction, is assigned to nuclei heavy enough 
to develop a central region of saturated nuclear matter. Via collective 
excitations of nuclei and magnetic multipole transitions, in particular the 
giant Gamow-Teller resonance, some of the nuclear Fermi-liquid parameters 
\cite{backman} can be determined from experimental data. Others are related to
bulk properties of nuclear matter, such as its compression modulus $K=k_{f0}^2 
\bar E''(k_{f0})$ or its (isospin) asymmetry energy $A(k_{f0})$. Various 
calculations of the nuclear Fermi-liquid parameters using Brueckner-Bethe 
theory and phenomenological nucleon-nucleon potentials have been performed in 
the past \cite{backman,sjoberg,dickh}. In ref.\cite{brown} the role of 
$\rho$-meson exchange for the spin-isospin parameter $G_0'$ has been 
emphasized. More recently, Schwenk et al. \cite{schwenk} have employed the 
renormalization group motivated universal low-momentum nucleon-nucleon 
potential $V_{\rm low-k}$ \cite{vlowk}. In their work the induced interaction 
generated by the particle-hole parquet diagrams plays a very important role. 
In contrast to this rather sophisticated approach, ref.\cite{kuckei} takes 
into account only the bare potential $V_{\rm low-k}$ or the in-medium
G-matrix. The resulting Fermi-liquid parameters are then not in good agreement
with empirical values (see Table 2 in ref.\cite{kuckei}). Moreover, saturation
of nuclear matter could not be obtained with the potential $V_{\rm low-k}$ in
the Brueckner-Hartree-Fock approximation. According to recent work of Bogner et
al. \cite{bsfn} the inclusion of the leading three-nucleon force from chiral 
effective field theory is essential in order to achieve reasonable saturation
properties in this approach.  

The purpose of the present paper is to explore the role of the long-range 
two-pion exchange for the nuclear Fermi-liquid parameters. We will restrict
ourselves to the isotropic part of the quasi-nucleon interaction for which
empirical information is best available \cite{backman,schwenk,speth}. To be
specific, consider the angle-averaged quasi-particle interaction $V_{\rm eff}$
of two nucleons on the Fermi sphere, $|\vec p_1|=|\vec p_2 | = k_f$. It is
irreducible in the direct particle-hole channel and it has the following form:
\begin{eqnarray} {\cal F}_0(k_f) &=& {1\over (4\pi)^2} \int d\Omega_1
d\Omega_2\, \langle  \vec p_1, \vec p_2 | V_{\rm eff}| \vec p_1,\vec p_2
\rangle \nonumber  \\ &=&  f_0(k_f) + g_0(k_f) \, \vec \sigma_1\cdot\vec
\sigma_2 + f'_0(k_f)\, \vec \tau_1 \cdot\vec \tau_2 + g'_0(k_f) \, \vec 
\sigma_1 \cdot\vec \sigma_2\, \vec \tau_1 \cdot \vec \tau_2 \,, \end{eqnarray}
where  $|\vec p_1,\vec p_2\rangle$ stands for a properly anti-symmetrized
two-nucleon state and $\vec \sigma_{1,2}$ and $\vec \tau_{1,2}$ denote the 
usual spin and isospin operators of the two nucleons. The density-dependent
functions $f_0(k_f),\, g_0(k_f),\, f'_0(k_f)$ and $g'_0(k_f)$ of dimension
fm$^2$ are the isotropic ($l=0$) Landau parameters. The spin- and 
isospin independent part $f_0(k_f)$ is alternatively given by the second
functional derivative of the energy density with respect to the occupation
density in momentum space \cite{sjoberg}. Because of this property the
Landau parameter $f_0(k_f)$ can be directly expressed in terms of the 
angle-integrated two-body and three-body kernels, ${\cal K}_2(p_1,p_2)$ and  
${\cal K}_3(p_1,p_2,p_3)$, introduced in ref.\cite{liquidgas} to facilitate 
the finite temperature calculation. For two-body and three-body contributions 
to $f_0(k_f)$ the following relations hold:  
\begin{equation} f_0(k_f)^{(\rm 2-body)} = {1 \over 8k_f^2} \, {\cal
K}_2(k_f,k_f) \,, \end{equation} 
\begin{equation} f_0(k_f)^{(\rm 3-body)} = {1 \over (4\pi k_f)^2} \int_0^{k_f}
dp\,p \Big[{\cal K}_3(p,k_f,k_f)+ {\cal K}_3(k_f,p,k_f)+{\cal K}_3(k_f,k_f,p)
\Big]\,, \end{equation} 
where the prefactors originate from the (convenient) normalization of the 
kernels ${\cal K}_{2,3}$ chosen in ref.\cite{liquidgas}. For an interpretation
of eqs.(2,3) note that in a diagrammatic language the second functional 
derivative is constructed by opening two nucleon lines of a closed in-medium 
diagram which represents the groundstate energy density. The relations  
eqs.(2,3) serve as a good check on our analytical one-loop calculation of the
isotropic Landau parameters $f_0(k_f),\, g_0(k_f),\, f'_0(k_f)$ and 
$g'_0(k_f)$ to be presented in the next section. 

\section{Diagrammatic calculation of isotropic Fermi-liquid parameters}
In this section we present analytical formulas for the density-dependent 
Fermi-liquid  parameters $f_0(k_f),\, g_0(k_f),\,f'_0(k_f)$ and $g'_0(k_f)$ as
derived from tree-level and one-loop pion-exchange diagrams. We give for each 
diagram only the final result omitting all technical details related to 
algebraic manipulations and solving elementary integrals. Diagrams can 
contribute via a direct term, where the ordering of $\vec p_1$ and $\vec p_2$ 
is the same in the initial and final state, and via a crossed term, where this
ordering is reversed in the final state. In the latter case the negative 
product of the spin- and isospin exchange operators $-(1+S)(1+T)/4$ has to be 
multiplied from the left. Here, we have introduced the (convenient) short hand 
notations, $S= \vec \sigma_1 \cdot\vec \sigma_2$ and $T= \vec \tau_1 \cdot\vec
\tau_2$, which we will use frequently in the following. The spin-spin and
isospin-isospin operators satisfy the relations $S^2 = 3-2S$ and $T^2 = 3-2T$.
We start now to enumerate the diagrammatic contributions to the angle-averaged
quasi-particle interaction ${\cal F}_0(k_f)$ defined in eq.(1).     

The crossed term from the (tree-level) one-pion exchange diagram in Fig.\,1 
reads:
\begin{equation}  {\cal F}_0(k_f) = {g_A^2 \over 48 f_\pi^2}(3-S)(3-T)  
\bigg[ 1 - {\ln(1+4u^2) \over 4u^2} \bigg] \bigg( 1 - {k_f^2 \over M^2} \bigg)
\,, \end{equation} 
with the useful abbreviation $u=k_f/m_\pi$, where $m_\pi =135\,$MeV stands for 
the (neutral) pion mass. As usual $f_\pi = 92.4\,$MeV denotes the weak pion
decay constant and we choose the value $g_A=1.3$ of the nucleon axial-vector
coupling constant in order to have a pion-nucleon coupling constant of 
$g_{\pi N} = g_A M/f_\pi = 13.2$. Furthermore, $M=939\,$MeV denotes the 
(average) nucleon mass. The relativistic correction factor $1-k_f^2/M^2$ in
eq.(4) is not part of the diagram, but it stems from the squared 
$M/E$-prefactor in the relativistic density of states, with $E^2=M^2+k_f^2$. 
Consistency with the two-body kernel ${\cal K}_2^{(1\pi)}$ in eq.(5) of
ref.\cite{liquidgas} requires the inclusion this relativistic correction
factor.        

\bigskip
\begin{figure}
\begin{center}
\includegraphics[scale=1.1,clip]{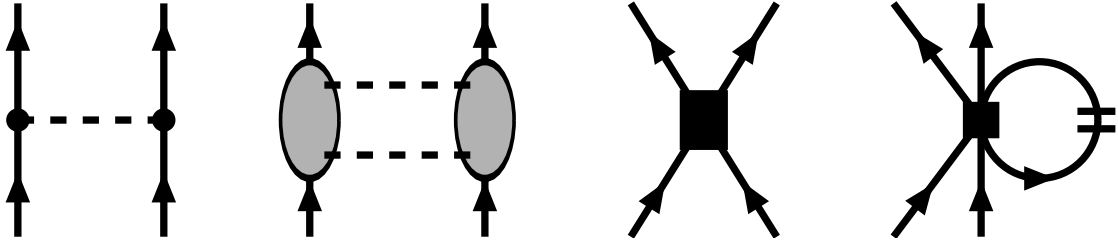}
\end{center}\vspace{-0.4cm}
\caption{In-medium scattering diagrams related to one-pion exchange, 
irreducible two-pion exchange, a two-body contact interaction, and a three-body
contact interaction.}    
\end{figure}

The left diagram in Fig.\,2 corresponds to iterated one-pion exchange. The
direct term of this one-loop graph gives rise to the following 
(spin-independent) contribution: 
\begin{equation}  {\cal F}_0(k_f) = {g_A^4 M m_\pi \over 512 \pi f_\pi^4 u^2}
(3-2T) \Big[ 16u \arctan 2u -5\ln(1+4u^2)\Big] \,, \end{equation}
whereas its crossed term contributes in the form:
\begin{eqnarray}  {\cal F}_0(k_f) &=& {g_A^4 M m_\pi \over 3\pi (4f_\pi)^4 u^2}
(3-5T) \bigg\{ {1\over 3}(3+S) \Big[4u^2-\ln(1+u^2) + (4u^3+6u) \arctan u 
\Big] \nonumber \\ && + (3-S) \bigg[ {1\over 2} \ln(1+4u^2) -2u 
\arctan 2u  +\int_0^u dx\,{\arctan 2x - \arctan x \over 1+2x^2} \bigg] \bigg\}
\,. \end{eqnarray}
Note the large scale enhancement factor $M$ which stems from an energy
denominator equal to the difference of small nucleon kinetic energies. The
expressions in eqs.(5,6) do not include the contribution of a 
linear divergence $\int_0^\infty dl\,1$ of the momentum-space loop integral. If
regularized by a cut-off scale $\Lambda$ as done in ref.\cite{nucmat} one gets
in addition the (constant) contribution:     
\begin{equation} {\cal F}_0(k_f) = {g_A^4 M \Lambda \over 128 \pi^2 f_\pi^4} 
(13 T-15 -3 S+5 S\,T) \,, \end{equation}
which is equivalent to that of a (momentum-independent) NN-contact coupling. 
It turns out that with respect to the constraints from the empirical nuclear 
matter binding energy $\bar E(k_{f0})$ and asymmetry energy $A(k_{f0})$, one 
adjusted cut-off $\Lambda$ is sufficient to represent the most general 
(i.e. two-parameter) contact-interaction at this order in the small momentum 
expansion. At higher orders there are of course more parameters to describe the
short-distance dynamics (see eqs.(22,40) below).\footnote{At N$^3$LO, 
corresponding to two-loop order in NN-scattering, there are in total 24
short-distance coefficients \cite{entem}.}       

\begin{figure}
\begin{center}
\includegraphics[scale=1.0,clip]{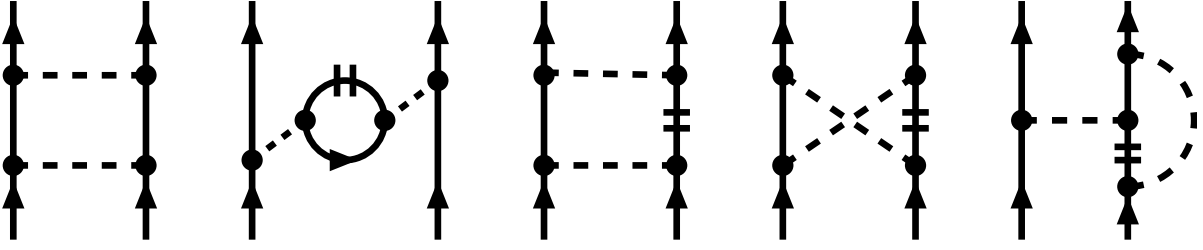}
\end{center}\vspace{-0.4cm}
\caption{Iterated one-pion exchange and medium modifications of one-pion
exchange and two-pion exchange. The short double-line symbolizes the
filled Fermi sea of nucleons, i.e. the medium insertion $-\theta(k_f -|\vec
p\,|)$ in the in-medium nucleon propagator \cite{nucmat}. Effectively, the
medium insertion sums up hole-propagation and the absence of 
particle-propagation below the Fermi surface $|\vec p\,| < k_f$. Mirror graphs
are not shown.}    
\end{figure}

The remaining four graphs in Fig.\,2 involve medium modifications as 
symbolized by the short double-line. The second diagram in Fig.\,2 together 
with its mirror partner can be interpreted as a one-pion exchange modified by 
the coupling of the pion to nucleon-hole states. From its crossed term we 
derive the following contribution:
\begin{equation} {\cal F}_0(k_f) = {g_A^4 M m_\pi \over 12 \pi^2 f_\pi^4 u^2}
(3-S)(3-T) \int_0^u  { dx\,x^4 \over (1+4x^2)^2} \bigg[ 2u x +(u^2-x^2) \ln{u+x
\over u-x} \bigg] \,. \end{equation}
Pauli blocking acts in the planar and crossed $2\pi$-exchange box diagrams in 
Fig.\,2 (and their mirror partners). The direct term from the planar box
diagram contributes in the form:  
 \begin{equation} {\cal F}_0(k_f) = {g_A^4 M m_\pi \over \pi^2 f_\pi^4 u^2}
(3-2T) \int_0^u { dx \,x^5 \over (1+4x^2)^2} \bigg[ u\ln{u+x \over 4(u-x)}  +x
\ln{u^2-x^2 \over x^2}    \bigg] \,, \end{equation}
whereas the direct term from the crossed box diagram with Pauli blocking reads:
\begin{equation} {\cal F}_0(k_f) = {g_A^4 M m_\pi \over \pi^2 f_\pi^4 u^2}
(3+2T) \int_0^u { dx \,x^5 \over (1+4x^2)^2} \bigg[ u\ln{4u^2 \over u^2-x^2}  
-x \ln{u+x \over u-x}    \bigg] \,. \end{equation}
Note that in both cases spin-dependent terms are absent. There is also the 
crossed term from the planar box with Pauli blocking. We split its contribution
to ${\cal F}_0(k_f)$ into a factorizable part:
\begin{equation} {\cal F}_0(k_f) = {g_A^4 M m_\pi \over 384\pi^2 f_\pi^4}
(3+S)(3-5T) \int_0^u dx \, \bigg[1+{x^2-u^2-1 \over 4u x} \ln{1+(u+x)^2 \over 
1+(u-x)^2}\bigg]^2 \,, \end{equation}
and a non-factorizable part:
\begin{eqnarray} {\cal F}_0(k_f) &=& {g_A^4 M m_\pi \over 48\pi^2 f_\pi^4u^2}
(3-S)(3-5T) \int_0^u  {dx \,x^3 \over 1+4x^2} \bigg[ u \ln{ 4(u-x)\over u+x}
\nonumber \\ && -x \ln{u^2-x^2 \over x^2} +\int_0^{u-x} {dy\, \over \sqrt{R}} 
\ln{u \sqrt{R} +(1-4x y)(x-y) \over u \sqrt{R} +(4x y-1)(x-y)} \bigg] \,, 
\end{eqnarray}
with the auxiliary polynomial $R=4u^2+(4x^2-1)(4y^2-1)$. These two pieces are 
distinguished by whether the (remaining) nucleon propagator in the denominator
can be canceled or not by terms from the product of $\pi N$-interaction 
vertices in the numerator. Finally, the
last diagram in Fig.\,2 (together with three mirror partners) represents a 
density-dependent vertex correction to the one-pion exchange. The
non-vanishing contribution comes from the crossed term and we split again into 
a factorizable part:
\begin{eqnarray}  {\cal F}_0(k_f) &=& {g_A^4 M m_\pi \over 3 \pi^2 (8f_\pi)^4 
u^5} (3-S)(3-T)\Big[4u^2-\ln(1+4u^2)\Big]\nonumber \\ && \times  
\Big[8u^4+4u^2-( 1+4u^2)\ln(1+4u^2)\Big] \,, \end{eqnarray}
and a non-factorizable part:
\begin{eqnarray} {\cal F}_0(k_f) &=& {g_A^4 M m_\pi \over 3\pi^2 (4f_\pi)^4u^2}
(3-S)(3-T) \int_0^u  dx \, \Big[ \ln(1+4x^2)-4x^2 \Big]  \nonumber \\ &&
\times \bigg\{  \ln{u+x \over u-x} + {1 \over \sqrt{1+4u^2-4x^2}} \ln { ( u
\sqrt{1+4u^2-4x^2} -x )^2 \over (1+4u^2)(u^2-x^2) } \bigg\} \,. \end{eqnarray}
As can be seen from the power of $m_\pi$ in their prefactors all contributions 
in eqs.(5,6,8-14) are of the same order in the small momentum expansion. The
(bare) one-pion exchange eq.(4) and the contact term eq.(7) are (formally) of
lowest order in the small momentum expansion.   

The p-wave $(l=1)$ Landau parameter $f_1(k_f)$, following $f_0(k_f)$ in the 
Legendre-polynomial expansion of the spin-isospin averaged quasi-nucleon 
interaction, is directly related to the slope of the (real) single-particle
potential  $U(p,k_f)$ at the Fermi surface \cite{landau,sjoberg}:
\begin{equation} f_1(k_f) = - {3\pi^2 \over 2k_f^2} \, {\partial U(p,k_f)
\over \partial p}\bigg|_{p= k_f} \,. \end{equation}
Its value at nuclear matter saturation density $\rho_0=2k^3_{f0}/3\pi^2$ 
determines the effective nucleon mass  $M^*$ via the relation:
\begin{equation} M^* = M \bigg[ 1- {k_{f0}^2 \over 2M^2}-{2Mk_{f0}\over 3\pi^2}
\,f_1(k_{f0}) \bigg]^{-1} \,. \end{equation} 
The second term $-k_{f0}^2/2M^2$ in the square brackets stems from the
relativistic correction $-p^4/8M^3$ to the kinetic energy. Although it is 
small, we keep this correction term for reasons of consistency with our earlier
works \cite{nucmat,deltamat}.  

\begin{figure}
\begin{center}
\includegraphics[scale=0.55,clip]{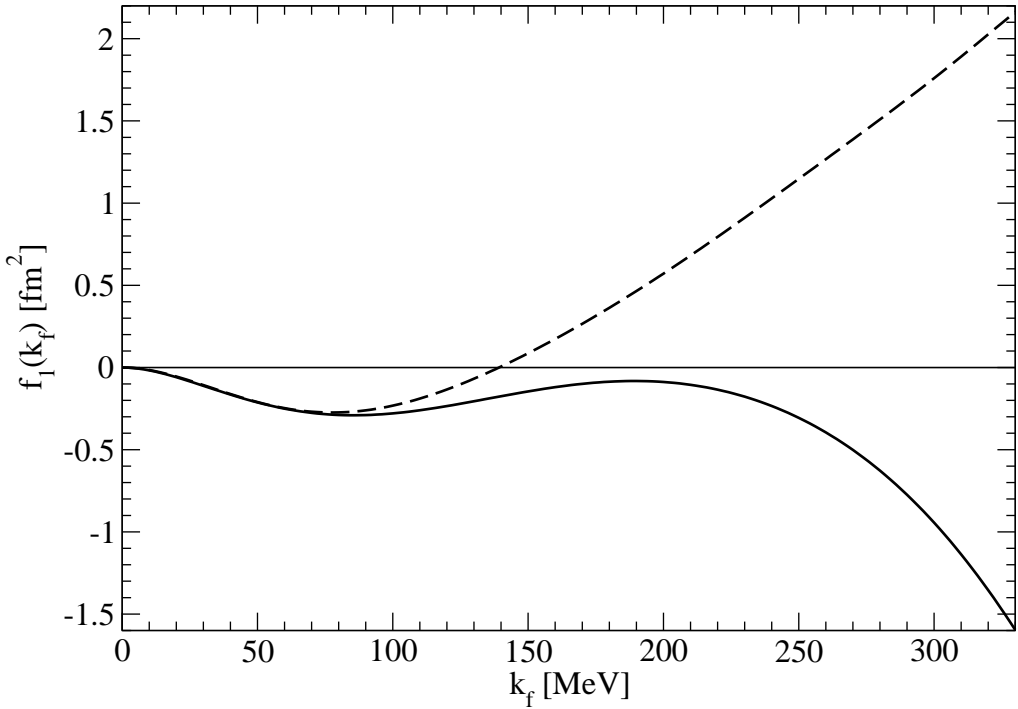}
\end{center}\vspace{-0.7cm}
\caption{The p-wave Landau parameter $f_1(k_f)$ as a function of the Fermi 
momentum $k_f$. The dashed line corresponds to the approximation to $1\pi$- 
and iterated $1\pi$-exchange. The full line includes in addition the 
contributions from irreducible  $2\pi$-exchange with no, single and double 
virtual $\Delta$-isobar excitation.}    
\end{figure}

Fig.\,3 shows the p-wave Landau parameter $f_1(k_f)$ as a function of the 
Fermi momentum $k_f$. The dashed curve corresponds to the approximation to
$1\pi$- and iterated $1\pi$-exchange \cite{nucmat,lutzcontra}. The full curve 
includes in addition the higher order contributions from irreducible 
$2\pi$-exchange with no, single and double virtual $\Delta$-isobar excitation 
\cite{deltamat}. One should note that all short-distance parameters related to 
momentum-independent NN-contact interactions drop out in the quantity $f_1(k_f
)$. The remaining possible short-distance contribution $f_1(k_f)^{(ct)}=-5
\pi^2  B_5 k_f^2/M^4$ from the momentum-dependent NN-contact interactions (see
eq.(9) in ref.\cite{deltamat}) plays numerically no role, since the adjustment
to the saturation curve $\bar E(k_f)$ and the potential depth $U(0,k_{f0})$ 
gave as an optimal value $B_5=0$ \cite{deltamat}. The curves in Fig.\,3
display therefore the pure long-range effects from chiral one- and two-pion
exchange. The upward bending of the dashed curve above $k_f \simeq 100\,$MeV
drives the p-wave Landau parameter $f_1(k_f)$ to a large positive value of
$f_1(2m_\pi)\simeq 1.4\,$fm$^2$ at saturation density which translates into an
unrealistically high effective nucleon mass of $M^*\simeq  2.9M$ at the Fermi
surface. As one can see from the full line in Fig.\,3 this wrong trend gets
with increasing Fermi momentum $k_f$ suppressed and finally reversed by the
inclusion of the chiral $\pi N\Delta$-dynamics. The negative value
$f_1(k_{f0}) \simeq -0.4\,$fm$^2$ at saturation density $k_{f0}= 261.6\,$MeV
\cite{deltamat} is compatible with the empirical values of $f_1(k_{f0})$, 
collected in Table\,1. As a consequence the effective nucleon mass at the 
Fermi surface takes on now a more realistic value: $M^* =0.88M$. For Fermi 
momenta as large as $k_{f0} \simeq 2m_\pi$ the higher order corrections from 
the $2\pi$-exchange with virtual $\Delta$-excitation turn out to be essential 
for good single-particle properties. This feature which is exhibited  here 
very clearly in Fig.\,3 is in agreement with the findings of 
ref.\cite{deltamat}.    
  
\begin{table}[hbt]
\begin{center}
\begin{tabular}{|c|ccccc|}
    \hline
   & $f_0$ [fm$^2$] & $f_1$ [fm$^2$] & $f_0'$ [fm$^2$] & $g_0$ [fm$^2$] & 
$g_0'$ [fm$^2$] \\  \hline   ref.\cite{backman}
& $-0.3\dots 0$ & $-0.75 \dots 0$ & $0.9\dots 1.5$ & small & $1.6\dots 1.7$ \\ 
 \hline  ref.\cite{speth} & $0.1$ & $-0.6$ & $0.7$ & $1.12$ & $1.41$ \\   
 \hline  ref.\cite{schwenk} & $-0.29$ & $-0.92$ & $0.77$ & $0.16\pm 0.3$ &
   $1.1 \pm 0.2$ \\
\hline  \end{tabular}
\end{center}
\end{table}
\vspace{-0.5cm}
Tab.1: Empirical values of the (dimensionful) Landau parameters at nuclear 
matter saturation density $k_{f0}=263\,$MeV in units of fm$^2$. The
dimensionless Landau parameters $F_0$, $F_1$, $F'_0$, $G_0$, $G'_0$ have been
divided by the density of states at the Fermi surface $N_0 = 2\pi^{-2}k_{f0}
M^*$ taking for the effective nucleon mass $M^* = 0.8 M$ \cite{speth} and $M^*
= 0.72 M$ \cite{schwenk}, respectively. In the first row we used $N_0=
1\,$fm$^{-2}$ (corresponding to  $M^* = 0.78 M$).   
\bigskip

Next, we discuss the results for the isotropic Landau parameters $f_0(k_f),\, 
g_0(k_f),\,f'_0(k_f)$ and $g'_0(k_f)$ in the approximation to $1\pi$- and 
iterated $1\pi$-exchange. By adjusting the single cut-off scale $\Lambda$ to 
the value $\Lambda = 611\,$MeV this approximation leads already to a good 
nuclear matter equation of state with the saturation point at $\rho_0 = 0.173\,
$fm$^{-3}$, $\bar E(k_{f0}) = -15.3\,$MeV, a nuclear matter compressibility
of $K = k^2_{f0}\bar E''(k_{f0}) = 252\,$MeV \cite{nucmat,lutzcontra}, and an
(isospin) asymmetry energy of $A(2m_\pi) = 38.9\,$MeV. Fig.\,4 shows the 
corresponding Landau parameters $f_0(k_f),\, g_0(k_f),\, f'_0(k_f)$ and 
$g'_0(k_f)$ (i.e. the summed contributions written in eqs.(4-14)) as a 
function of the Fermi momentum $k_f$. The strong density dependence of 
$f_0(k_f)$ including a sign change in the vicinity of saturation density 
$k_{f0}$ is a generic feature which is also shared by many other 
calculations \cite{backman}. The other three Fermi-liquid parameters $g_0(k_f),
\,f'_0(k_f)$ and $g'_0(k_f)$ vary much less with the density $\rho = 2k_f^3/3
\pi^2$ (or the Fermi momentum $k_f$). We read off from Fig.\,4 at saturation 
density: $f_0(2m_\pi) =0.55\,$fm$^2$, $g_0(2m_\pi) = -1.44 \,$fm$^2$, 
$f'_0(2m_\pi)= 2.01\,$fm$^2$, and  $g'_0(2m_\pi)= 1.14\,$fm$^2$. The negative 
value of $g_0(2m_\pi)=-1.44\,$fm$^2$ reflects the spin-instability of nuclear 
matter in this approximation, as discussed recently in ref.\cite{spinstab}. 
Although they lie outside their empirical ranges (see Table\,1) the predicted 
values $f_0(k_{f0}) =0.55\,$fm$^2$ and  $f'_0(k_{f0})= 2.01\,$fm$^2$ do obey 
the Landau relations \cite{landau,sjoberg} to the nuclear matter 
compressibility:
\begin{equation} K = 3k_{f0}^2 \bigg[ {1\over M^*}+{2k_{f0}\over \pi^2} \, 
f_0(k_{f0})\bigg]\,, \end{equation}
and to the (isospin) asymmetry energy:
\begin{equation} A(k_{f0}) = k_{f0}^2 \bigg[{1\over 6M^*}+{k_{f0}\over 3\pi^2} 
\, f_0'(k_{f0})\bigg]\,.\end{equation}
The discrepancy comes of course from the much too high effective nucleon mass
$M^* \simeq  2.9M$ in this approximation. The two spin-dependent Fermi-liquid 
parameters $g_0(k_{f0})$ and $g_0'(k_{f0})$ are connected by relations 
analogous to eq.(18) to the spin-asymmetry energy $S(k_{f0})$ and the 
spin-isospin asymmetry energy $J(k_{f0})$ \cite{spinstab}. We point out that
in our calculation these (four) consistency relations hold with a numerical 
precision of one permille and better. It is also interesting to look at 
individual contributions. For example, the spin-isospin interaction strength 
$g'_0(2m_\pi)$ decomposes as  $g'_0(2m_\pi) = (0.13+0.99+0.02)\,$fm$^2= 
1.14\,$fm$^2$ into contributions from (static) $1\pi$-exchange eq.(4), 
iterated  $1\pi$-exchange eqs.(6,7), and medium modified terms. Interestingly,
the medium modified terms eqs.(8,11-14) sum up to a negligibly small net
contribution. The important role of iterated  $1\pi$-exchange for the 
spin-isospin Fermi-liquid parameter $g'_0(k_f)$ has been stressed earlier in
ref.\cite{backman}. The new element here is that this contribution is now
evaluated consistently in a framework which leads to realistic nuclear matter
binding and saturation. The predicted value $g'_0(2m_\pi) = 1.14\,$fm$^2$ of 
the spin-isospin Fermi-liquid parameter is compatible with existing empirical 
values (see Table\,1). Nevertheless, it is clear from Fig.\,3 that higher 
order corrections from $2\pi$-exchange with virtual $\Delta$-isobar excitation
are essential in order to obtain good single-particle properties. Therefore we
will now turn to the contributions of the chiral $\pi N\Delta$-dynamics to the
isotropic Fermi-liquid parameters $f_0(k_f),\, g_0(k_f),\,f'_0(k_f)$ and
$g'_0(k_f)$.

\begin{figure}
\begin{center}
\includegraphics[scale=0.75]{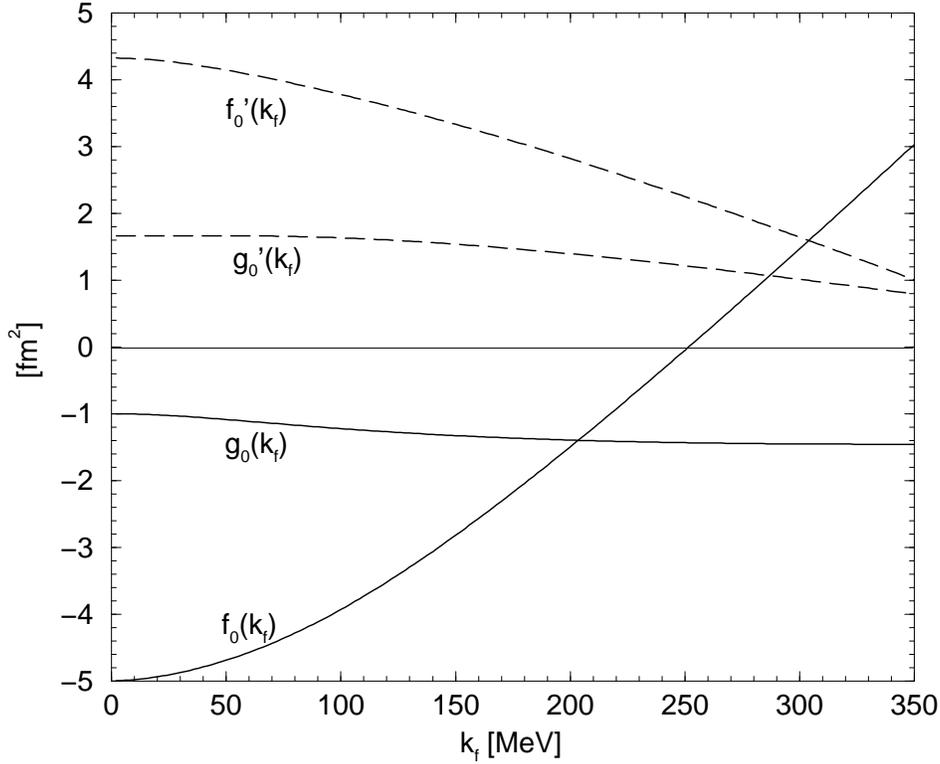}
\end{center}\vspace{-0.7cm}
\caption{The isotropic Landau parameters $f_0(k_f),\, g_0(k_f),\,f'_0(k_f)$ 
and $g'_0(k_f)$ as a function of the Fermi momentum $k_f$. Only contributions 
from single and iterated pion exchange are included.}    
\end{figure}

\subsection{Two-pion exchange with virtual $\Delta$-isobar excitation} 
The two-pion exchange nucleon-nucleon potential is symbolized by the second
diagram in Fig.\,1. We separate regularization dependent short-distance parts 
from the unique long-range terms with the help of a twice-subtracted
dispersion relation. Applying this procedure to the crossed term of the 
(irreducible) $2\pi$-exchange diagram in Fig.\,1 we obtain a contribution to
${\cal F}_0(k_f)$ of the form:
\begin{eqnarray} {\cal F}_0(k_f) &=& {1 \over \pi} \int_{2m_\pi}^\infty d\mu
\bigg[ {k_f^2 \over \mu^3} - {1\over 2\mu} +{\mu \over 8 k_f^2} \ln\bigg( 1 +
{4k_f^2 \over \mu^2} \bigg) \bigg] \bigg\{ {\rm Im}\Big(V_C+3W_C+2\mu^2 V_T+
6\mu^2W_T\Big) \nonumber \\ && + S \,{\rm Im}\Big(V_C+3W_C-{2\over 3}\mu^2 
V_T-2\mu^2 W_T\Big) + T \,{\rm Im}\Big(V_C-W_C+2\mu^2 V_T-2\mu^2 W_T \Big)
\nonumber \\ && + S \, T \, {\rm Im}\Big(V_C-W_C-{2\over 3}\mu^2 V_T+{2\over 
3}\mu^2 W_T\Big) \bigg\}\,,   \end{eqnarray}
where Im$V_C$, Im$W_C$, Im$V_T$ and Im$W_T$ are the spectral functions of the
isoscalar and isovector central and tensor NN-amplitudes, respectively. 
Explicit expressions of these imaginary parts for the contributions of the 
triangle diagram with single $\Delta$-excitation and the box diagrams with 
single and double $\Delta$-excitation can be easily constructed from the 
analytical formulas given in section 3 of ref.\cite{gerst}. The $\mu$- and 
$k_f$-dependent weighting function in eq.(19) takes care that at low and
moderate Fermi momenta this spectral integral is dominated by low invariant
$\pi\pi$-masses $2m_\pi< \mu <1\,$GeV. The contributions to ${\cal F}_0(k_f)$ 
from irreducible $2\pi$-exchange with only nucleon intermediate states can 
also be cast into the form eq.(19). The corresponding non-vanishing spectral 
functions read \cite{nnpap}:
\begin{equation} {\rm Im}W_C(i\mu) = {\sqrt{\mu^2-4m_\pi^2} \over 3\pi 
\mu (4f_\pi)^4} \bigg[ 4m_\pi^2(1+4g_A^2-5g_A^4) +\mu^2(23g_A^4-10g_A^2-1) + 
{48 g_A^4 m_\pi^4 \over \mu^2-4m_\pi^2} \bigg] \,, \end{equation}
\begin{equation} {\rm Im}V_T(i\mu) = - {6 g_A^4 \sqrt{\mu^2-4m_\pi^2} \over 
\pi  \mu (4f_\pi)^4}\,. \end{equation}
The subtraction constants which come along with the twice-subtracted
dispersion relation representation of the $2\pi$-exchange nucleon-nucleon 
potential effectively parameterize a two-body contact interaction (see third 
diagram in Fig.\,1). The direct and crossed term from the most general 
two-body contact interaction up to order-$p^2$ give rise together to the 
contribution:
\begin{eqnarray} {\cal F}_0(k_f) &=& {\pi^2 \over M^2} \Big\{ 3B_3
+S\,(B_3-6B_{n,3}) +T\,(6B_{n,3}-3B_3) -S\,T \, B_3 \Big\} \nonumber \\ &&
+{5\pi^2 k_f^2\over M^4}\Big\{B_5 +T\, (2B_{n,5}-B_5) +S\, B_5^{\sigma}
+ S\, T\,B_5^{\sigma\tau}\Big\}\,.  \end{eqnarray}
Note that the $k_f$-independent term in eq.(22) involves only two independent 
subtraction constants. This is a consequence of the Pauli exclusion principle
(or the Fierz-antisymmetric nature of the two-body contact interaction). The 
four (dimensionless) parameters $B_3 = -7.99$, $B_{n,3} = -0.95$, $B_5 =0$ and 
$B_{n,5} = -3.58$ have been adjusted in ref.\cite{deltamat} to empirical 
nuclear nuclear properties (such as the maximal binding energy per particle
$-\bar E(k_{f0}) = 16\,$MeV and the isospin asymmetry energy $A(k_{f0}) =
34\,$MeV). We note as an aside that the constant contribution eq.(7) linear in 
the cut-off $\Lambda$ is of course now not counted extra since the parameters 
$B_3$ and $B_{n,3}$ collect all such possible terms. The remaining two 
parameters $B_5^{\sigma}$ and $B_5^{\sigma\tau}$ in eq.(22) are a priori not 
constrained by any groundstate properties of spin-saturated nuclear matter. 
We have also checked that the usual symmetries of the NN-interaction (isospin
and Galilean invariance and Fierz-antisymmetry) do not imply a (linear)
relation between the four parameters $B_5,\,B_{n,5},\,B_5^{\sigma}$ and
$B_5^{\sigma\tau}$.  

In order to reduce a too strongly repulsive $\rho^2$-term in the energy per
particle $\bar E(k_f)$ a three-body contact interaction has been introduced in 
ref.\cite{deltamat}. The corresponding contribution to the quasi-particle 
interaction (represented by the last diagram in Fig.\,1) is of the form:   
\begin{equation} {\cal F}_0(k_f) =  {g_A^4 \zeta k_f^3 \over 8\pi^2 \Delta
f_\pi^4 }(3-S-T-S\, T) \,, \end{equation} 
with $\Delta=293\,$MeV the delta-nucleon mass splitting and the numerical 
parameter $\zeta = -3/4$. Here we have taken over from ref.\cite{deltamat} 
the parameterization of three-body contact-coupling strength. It is important 
to note here that the Pauli exclusion principle (together with isospin and
rotational invariance) allows only for one single momentum-independent 
three-nucleon contact-coupling $\sim(\zeta g_A^4/\Delta f_\pi^4)\,(\bar 
NN)^3$. 
  
\bigskip
\begin{figure}
\begin{center}
\includegraphics[scale=1.1,clip]{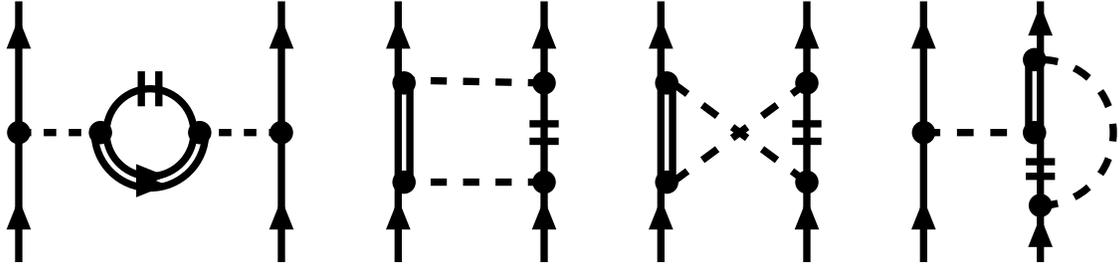}
\end{center}\vspace{-0.5cm}
\caption{Medium modifications of one-pion exchange and two-pion exchange with
 excitation of virtual $\Delta(1232)$-isobars. Mirror graphs are not shown.}
\end{figure}

Fig.\,5 shows additional pion-exchange diagrams with (single) virtual 
$\Delta(1232)$-isobar excitation involving medium modifications. The first 
diagram in Fig.\,5 can be interpreted as a one-pion exchange modified by the 
coupling to delta-hole states. Its crossed term gives rise to following the 
contribution:
\begin{equation} {\cal F}_0(k_f) =  {g_A^4 m_\pi^3 u  \over 144\pi^2 \Delta
f_\pi^4 }(3-S)(3-T)  \bigg[ 2u^2 +{2u^2 \over 1+4u^2} - \ln(1+4u^2) \bigg] \,,
\end{equation}  
with $u = k_f/m_\pi$, and we have already inserted the empirically well 
satisfied relation $g_{\pi N\Delta} = 3g_{\pi N}/\sqrt{2}$ for the $\pi N 
\Delta$-coupling constant. The delta propagator shows up in this expression
merely via the reciprocal mass splitting $\Delta =293\,$MeV. Pauli blocking 
acts in the second and third (planar and crossed) $2\pi$-exchange box diagrams
with $\Delta$-excitation. The summed contribution of their direct terms is 
spin- and isospin independent:
\begin{equation} {\cal F}_0(k_f) =  {g_A^4 m_\pi^3   \over 16\pi^2 \Delta
f_\pi^4 }\bigg[ 4u^3 -12u+15 \arctan 2u -{9\over 2u}\ln(1+4u^2) \bigg] \,.
\end{equation}
On the other hand the summed contribution of their crossed terms can be
written in the form:  
\begin{equation} {\cal F}_0(k_f) =  -{g_A^4 m_\pi^3   \over 64\pi^2 \Delta
f_\pi^4 }\int_0^u dx \, \bigg\{ 2X^2+Y^2 +{S+T \over 3} \, (2X^2+7Y^2)
 +{S\, T \over 9} \, (10X^2+17Y^2) \bigg\} \,, \end{equation}
with the two auxiliary functions:
\begin{equation} X = 2 x -{1\over 2u} \ln{1+(u+x)^2 \over 1+(u-x)^2} \,,
\end{equation}
\begin{equation} Y = {5x^2-3u^2-3\over 4 x}+{4x^2+3(1+u^2-x^2)^2 \over 16 u
 x^2}\ln{1+(u+x)^2 \over 1+(u-x)^2} \,. \end{equation} 
Finally, the last diagram in Fig.\,5 (together with three mirror partners) 
represents a density-dependent vertex correction to the one-pion exchange 
involving virtual $\Delta$-excitation. The resulting contribution  to ${\cal
F}_0(k_f)$ from the crossed term reads:
\begin{eqnarray}  {\cal F}_0(k_f) &=&  {g_A^4 m_\pi^3   \over 9\pi^2 \Delta (4
f_\pi)^4}(3-S)(3-T) \bigg\{ {8\over u^2}\Big[ \ln(1+4u^2)-4u^2 \Big] \arctan 2u
\nonumber \\  && +35u -{68u^3\over 3}-{3+4u^2\over 4u^3 }- {3+16u^2 +144 u^4
\over 64 u^7 } \ln^2(1+4u^2) \nonumber \\  &&+ {9+30u^2 -12 u^4 +112
u^6 \over 24 u^5 }  \ln(1+4u^2) \bigg\} \,. \end{eqnarray}

\begin{figure}
\begin{center}
\includegraphics[scale=0.7]{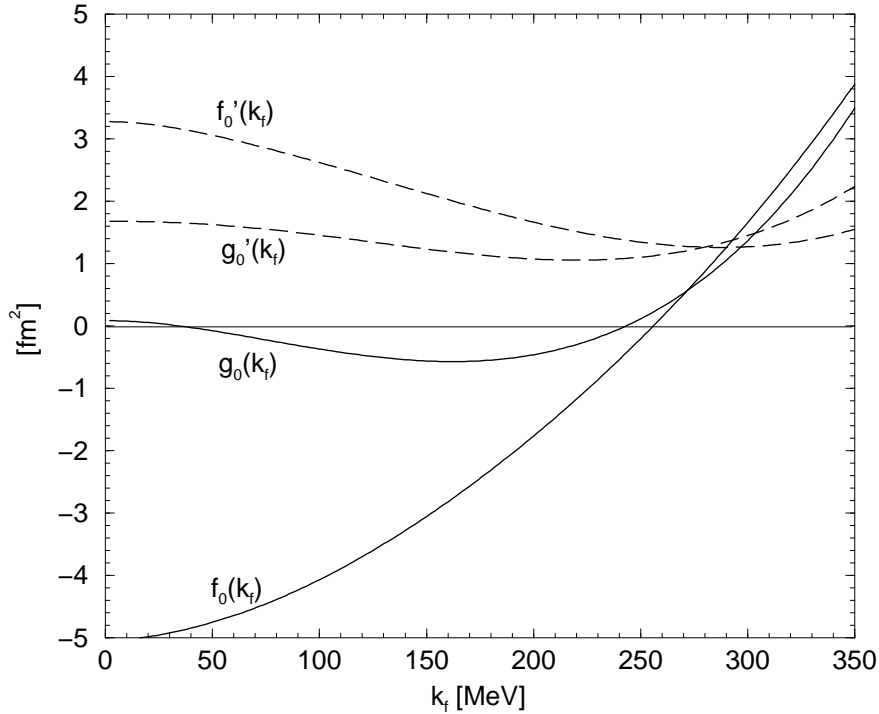}
\end{center}\vspace{-0.9cm}
\caption{The isotropic Landau parameters $f_0(k_f),\,g_0(k_f),\,f'_0(k_f)$ and 
$g'_0(k_f)$ as a function of the Fermi momentum $k_f$. Contributions from 
two-pion exchange with excitation of virtual $\Delta(1232)$-isobars are
included in addition. The (spin-dependent) short-distance parameters have been
set to the values $B_5^\sigma = - 14$ and $B_5^{\sigma\tau} = - 11$.}     
\end{figure}

Fig.\,6 shows again the isotropic Fermi-liquid parameters $f_0(k_f),\,g_0(k_f)
,\,f'_0(k_f)$ and $g'_0(k_f)$ as a function of the Fermi momentum $k_f$,
including now the new contributions written down in eqs.(19-29). One observes 
that the curve for $f_0(k_f)$ has approximately the same strong density
dependence as before (compare with Fig.\,4). The value $f_0(k_{f0})= 0.20\,
$fm$^2$ at saturation density ($k_{f0}= 261.6\,$MeV \cite{deltamat}) lies now
much closer to its empirical range (see Table\,1). The remaining 
overestimation of $f_0(k_{f0})$ is a consequence of the somewhat too high
nuclear matter compressibility $K= 304\,$MeV obtained in ref.\cite{deltamat}. 
The curves for $f_0(k_f)$ in Figs.\,4,6 are also quite instructive since they
reveal that the smallness of the empirical $f_0(k_{f0})$ simply results from a
zero-crossing of $f_0(k_f)$ in the near vicinity of nuclear matter saturation
density. The isospin dependent Fermi-liquid parameter $f'_0(k_f)$ varies now
less with density (compare Fig.\,6 with Fig.\,4) and it develops even a
minimum at $k_f\simeq 290\,$MeV. The resulting value at saturation density
$f_0'(k_{f0})=1.30\,$fm$^2$ lies within the empirically allowed range (see
first row in Table\,1). We note again that the consistency relations 
eqs.(17,18) hold with high numerical precision in our calculation. However, 
there are also the Pauli principle sum rules (see appendix A in
ref.\cite{backman}). Due to their non-linear character these sum rules are 
most likely violated in the present perturbative approach. 

Next, we discuss the spin-dependent Fermi-liquid parameters $g_0(k_f)$ and 
$g'_0(k_f)$. If one leaves out the $k_f^2$-contributions proportional to the 
short-distance parameters $B_5^{\sigma}$ and $B_5^{\sigma\tau}$ completely
(i.e. setting $B_5^{\sigma}= B_5^{\sigma\tau}=0$) one gets considerably too 
high values at saturation density, namely $g_0(k_{f0})= 2.7\,$fm$^2$ and
$g'_0(k_{f0})= 3.0\,$fm$^2$. Relatively large contributions come from the
twice-subtracted $2\pi$-exchange potential eq.(19): $(2.33\,S + 1.07\,S T)\,
$fm$^2$, and the three-body term eq.(23): $(0.89\,S + 0.89\,S T)\,$fm$^2$. The 
additional short-range dynamics encoded in the parameters $B_5^{\sigma}$ and 
$B_5^{\sigma\tau}$ is therefore important in the spin-dependent channels. This 
is in line with ref.\cite{brown} where the role of $\rho$-meson exchange has 
been emphasized. In order to get an estimate of the short-distance parameters 
$B_5^{\sigma}$ and $B_5^{\sigma\tau}$ we bring into play the complete set of 
four-nucleon contact-couplings written down in eqs.(3,4) of ref.\cite{evgeni}.
This set represents the most general short-range NN-interaction quadratic in 
momenta and it involves seven low-energy constants $C_1, \dots, C_7$. After 
computing the spin-dependent part of ${\cal F}_0(k_f)$ from the direct and
crossed term of the corresponding contact-potential we find:      
\begin{eqnarray} B_5^{\sigma} &=& {M^4 \over 30 \pi^2}(-3C_1+3C_3+3C_4+C_6+C_7)
\nonumber \\ &=&   {M^4 \over 320 \pi^3} \Big[-6C(^1\!S_0)+2 C(^3\!S_1)-
3C(^1\!P_1)+C(^3P_0) +3C(^3\!P_1)+5C(^3\!P_2)\Big]\,, \end{eqnarray}
\begin{eqnarray} B_5^{\sigma\tau} &=& {M^4 \over 30 \pi^2}(-3C_1+3C_3+C_6)
\nonumber \\ &=& {M^4 \over 960 \pi^3} \Big[-6C(^1\!S_0)-6 C(^3\!S_1)+
9C(^1\!P_1)+C(^3\!P_0)+3C(^3\!P_1)+5C(^3\!P_2)\Big]\,.\end{eqnarray}
In that form we obtain from the entries of Table IV in
ref.\cite{evgeni} (corresponding to various realistic NN-potentials) the
ranges $B_5^{\sigma} =-7.4\dots-6.8 $ and $B_5^{\sigma\tau}= -4.0 \dots-2.3$.
It is gratifying that these estimates lead to sizeable negative values of the 
short-distance parameters $B_5^{\sigma}$ and $B_5^{\sigma\tau}$, which are 
actually needed in order to reduce the far too high values of $g_0(k_{f0})= 
2.7\,$fm$^2$ and $g'_0(k_{f0})= 3.0\,$fm$^2$ mentioned before. The curves in 
Fig.\,6 for the spin-dependent Fermi-liquid parameters $g_0(k_f)$ and 
$g'_0(k_f)$ have been calculated with $B_5^{\sigma}= -14$ and $B_5^{\sigma
\tau} =-11$. This choice reproduces the empirical values $g_0(k_{f0})=0.34
\,$fm$^2$ and $g'_0(k_{f0})=1.15\,$fm$^2$ of ref.\cite{schwenk}. Another 
welcome feature is that $g_0(k_f)$ does now not fall below $-0.57\,$fm$^2$ and
therefore the stability condition $2\pi^{-2} M^*k_f\,g_0(k_f)> -1$ 
\cite{landau,sjoberg} is satisfied for all relevant densities. One can also 
see from Table\,1 that the empirical value of $g_0(k_{f0})$ has a large
uncertainty. In comparison to the lower ends of the foregoing estimate from 
realistic NN-potentials sizeable enhancement factors of about 1.9 and 2.8 need
to  be applied to the short-distance parameters $B_5^{\sigma}$ and $B_5^{\sigma
\tau}$. Obviously, our treatment of the short-range dynamics leaves here large
errors. It should also be mentioned that most realistic NN-potentials (with 
the exception of $V_{\rm low-k}$ \cite{bsfn}) require Brueckner resummation to
yield meaningful results in nuclear matter.   

The present calculation has still another shortcoming. According to earlier 
works \cite{babu,sjoberg,schwenk,wambach} the induced interaction
(which sums planar exchange particle-hole diagrams to infinite order) is an
important contribution to the quasi-particle interaction in nuclear matter. 
Some of the one-loop diagrams in Figs.\,2,5 do generate
leading pion-exchange contributions to the induced interaction, but all the
higher order iterations are consistently dropped here. In view of the good
nuclear matter and single-particle properties obtained in the present
perturbative framework one could argue that their effects in the
spin-independent channels ($f_{0,1}(k_f)$ and $f_0'(k_f)$) are hidden in the
adjusted values of the short-distance parameters. In the spin-dependent
channels ($g_0(k_f)$ and $g_0'(k_f)$) the missing higher order iterations
subsumed in the induced interaction may be the reason for the large mismatch
remaining from the estimates of $B_5^{\sigma}$ and $B_5^{\sigma\tau}$ from
realistic NN-potentials. It should also be mentioned that the estimate $B_5
\simeq 7.8$  from these NN-potentials is not consistent with the optimal value
$B_5 = 0$ of ref.\cite{deltamat}.

\section{Tensor interaction}
In this section we investigate along the same lines as in section 2 the tensor
part of the interaction of quasi-nucleons on the Fermi surface. The 
non-central (relative) tensor interaction has the following form \cite{dabr}: 
\begin{equation} {1\over 4k_f^2} \Big(3\,\vec \sigma_1 \cdot\vec q
\,\vec \sigma_2\cdot \vec q -\vec \sigma_1 \cdot\vec \sigma_2\, {\vec q}^{\,2}
\Big) \sum_{l=0}^\infty \Big[ h_l(k_f) +  h'_l(k_f)\, \vec \tau_1 \cdot\vec 
\tau_2 \Big]\,P_l(\cos\vartheta)\,, \end{equation}
with $\vec q = \vec p_1-\vec p_2$ the difference between two (incoming or
out-going) nucleon momentum vectors on the Fermi sphere $|\vec p_1|=|\vec p_2|
=k_f$. The angular dependence of the tensor interaction strength is
represented in eq.(32) by a series in Legendre-polynomials $P_l(\cos\vartheta
)$ of $\cos \vartheta = \hat p_1 \cdot \hat p_2$. Again, we will restrict
ourselves here to the (analytical) calculation to the leading ($l=0)$ 
Fermi-liquid parameters $h_0(k_f)$ and $h'_0(k_f)$ belonging to the tensor 
interaction. A first simplification arises
from the observation that only crossed terms of pion-exchange diagrams can 
contribute to the tensor interaction. The different ordering of $\vec p_1$ and
$\vec p_2$ in the initial and final state is a necessary condition for the
spin-operators $\vec \sigma_{1,2}$ and the momentum transfer $\vec q= \vec
p_1-\vec p_2$ to survive in the diagrammatic amplitude. The contribution of
the (bare) one-pion exchange is well known \cite{backman,brown}:
\begin{equation} h_0(k_f) +  T\, h'_0(k_f) = {g_A^2 \over 24 f_\pi^2} (3-T)
\ln(1+4u^2) \bigg( 1 - {k_f^2 \over M^2} \bigg) \,, \end{equation}
with $u= k_f/m_\pi$, and we have included the same relativistic correction 
factor as in eq.(4). The contribution from iterated one-pion exchange (see 
left diagram in Fig.\,2) has the form:
\begin{eqnarray}   h_0(k_f) +  T\, h'_0(k_f) &=& {g_A^4 M m_\pi \over 3\pi 
(4f_\pi)^4}(3-5T) \bigg\{ 2 + \ln{1+u^2\over 1+4u^2} -2u \arctan u \nonumber 
\\ &&  -{1\over u}\arctan 2u  + 2\int_0^u dx\,{\arctan 2x - \arctan x \over
1+2x^2} \bigg\} \,, \end{eqnarray}  
and the next diagram where the pion couples to nucleon-hole states leads to 
the expression:
\begin{equation} h_0(k_f) +  T\, h'_0(k_f) = {g_A^4 M m_\pi \over 6 \pi^2 
f_\pi^4 }(3-T) \int_0^u  { dx\,x^2 \over (1+4x^2)^2} \bigg[ 2u x +(u^2-x^2)
\ln{u+x \over u-x} \bigg] \,. \end{equation} 
More involved is the evaluation of the tensor component from the planar 
$2\pi$-exchange box diagram with Pauli blocking (see third diagram in 
Fig.\,2). We end up with the following triple integral representation:
\begin{eqnarray}   h_0(k_f) +  T\, h'_0(k_f) &=& {g_A^4 M k_f \over 96
\pi^2 f_\pi^4}(3-5T) \int_0^1 \!dx \int_{-1}^1\!dy \int_{-1}^1\!dz \, {x^2 
\over (u^{-2}+A)(u^{-2}+B) } \nonumber  \\ && \times  \bigg\{ {1\over 2}+{x^2
|y+z| \over A+B-4} - {(1-x^2)^2 \, {\rm sign}[(x-y)(x-z)] \,\theta(W)
\over  (A+B-4) \sqrt{W} } \bigg\} \,, \end{eqnarray} 
with the auxiliary polynomials $A= 1+x^2-2x y $, $B= 1+x^2-2x z$ and $W= 
(x-y)^2(x-z)^2-(1-y^2)(1-z^2)$. Finally, we have the last pion-exchange 
diagram in Fig.\,2 with a density-dependent vertex correction. We split its
contribution to the tensor Fermi liquid parameters into a factorizable
part:
\begin{equation}  h_0(k_f) +  T\, h'_0(k_f)= {g_A^4 M m_\pi \over 6 \pi^2 
(4f_\pi)^4 u^3}(3-T) \ln(1+4u^2) \Big[8u^4+4u^2-( 1+4u^2)\ln(1+4u^2)\Big] \,, 
\end{equation}
and a nonfactorizable part:
\begin{equation} h_0(k_f) +  T\, h'_0(k_f)= {g_A^4 M m_\pi \over 96\pi^2
f_\pi^4} (3-T) \int_0^u  dx \, {\ln(1+4x^2)-4x^2 \over \sqrt{1+4u^2-4x^2}} 
\ln { ( u\sqrt{1+4u^2-4x^2} +x )^2 \over (1+4u^2)(u^2-x^2) }\,. \end{equation}

\begin{figure}
\begin{center}
\includegraphics[scale=0.75]{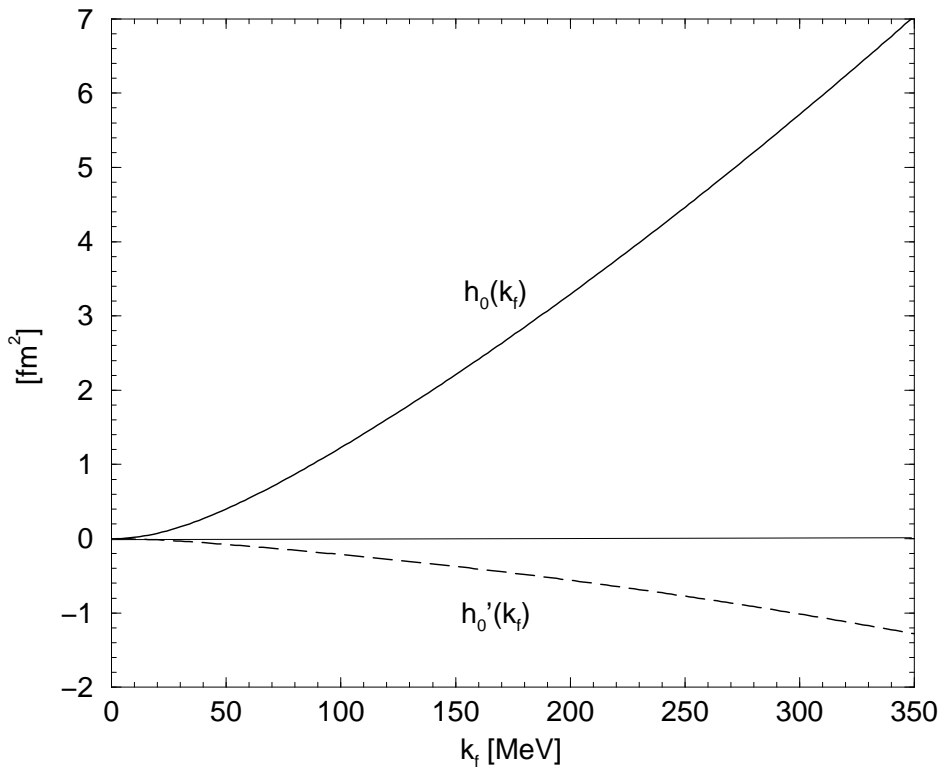}
\end{center}
\vspace{-0.7cm}
\caption{The Landau parameters $h_0(k_f)$ and $h'_0(k_f)$ related to the
tensor interaction as a function of the Fermi momentum $k_f$. Only 
contributions from single and iterated pion exchange are included.}    
\end{figure}

Fig.\,7 shows the summed contributions eqs.(33-38) to the tensor Landau
parameters $h_0(k_f)$ and $h'_0(k_f)$ as a function of the Fermi momentum
$k_f$. We have dropped the small relativistic $1/M^2$-correction to the
$1\pi$-exchange term eq.(33), since compared to the other contributions it is 
of higher order in the small momentum expansion. The solid and dashed line in
Fig.\,7 correspond to the isoscalar $h_0(k_f)$ and isovector $h_0'(k_f)$
tensor interaction strength, respectively. Both curves start quadratically in 
$k_f$ and above $k_f \simeq 100\,$MeV a more linear behavior takes over. At
nuclear matter saturation density $k_{f0}=263\,$MeV we obtain the values 
$h_0(k_{f0})= 4.78\,$fm$^2$ and $h_0'(k_{f0})=-0.83\,$fm$^2$. These numbers 
are to be compared with the leading tensor interaction from (static) 
$1\pi$-exchange which gives $h_0(k_{f0})^{1\pi}= 2.68\,$fm$^2$ and $h_0'
(k_{f0})^{1\pi} =-0.89\,$fm$^2$. Quite surprisingly, the (parameterfree)
$2\pi$-exchange corrections lead to almost a doubling of the isoscalar tensor
strength $h_0(k_{f0})$, whereas the isovector tensor strength $h_0'(k_{f0})$ is
much less affected. It is also instructive to see the individual contributions
from the iterated $1\pi$-exchange: $(-0.90+ 1.50\,T)\,$fm$^2$, the pion
coupling to nucleon-hole states: $(3.75-1.25\,T)\,$fm$^2$, the planar box with
Pauli blocking: $(0.33-0.55\,T)\,$fm$^2$, and the vertex correction: $(-1.08+
0.36\,T)\,$fm$^2$. In comparison to these results previous works 
\cite{brown,dabr} have found much smaller corrections to the one-pion exchange
tensor interaction. The difference comes of course from our explicit treatment
of the iterated pion-exchange (proportional to the large nucleon mass $M$) 
which is known to generate a sizeable tensor interaction (see eq.(32) in
ref.\cite{nnpap}). 

Next, we are interested in the effects of the chiral $\pi N\Delta$-dynamics
on the tensor interaction at the Fermi surface. The long-range part of the
corresponding $2\pi$-exchange NN-potential (symbolized by the second diagram 
in Fig.\,1) leads to the following contribution:                
\begin{equation}  h_0(k_f) +  T\, h'_0(k_f) = {1 \over 3\pi} 
\int_{2m_\pi}^\infty d\mu \bigg[ \mu \ln\bigg( 1 +{4k_f^2 \over \mu^2} \bigg)-
 {4k_f^2 \over \mu}\bigg] \Big\{ {\rm Im}(V_T+3W_T) +T \,  {\rm Im}(V_T-W_T) 
\Big\} \,, \end{equation}
where Im$V_T$ and Im$W_T$ are the spectral functions of the isoscalar and 
isovector tensor NN-amplitudes \cite{gerst}. The short-range pieces are again 
encoded in two new subtraction constants:  
\begin{equation}  h_0(k_f) +  T\, h'_0(k_f) = {k_f^2 \over 32\pi} \bigg\{ 
2 \sqrt{2} (1-T) C(\epsilon_1) +{1\over 3}(3+T) \Big[ 3C(^3\!P_1)-2 C(^3\!P_0)
- C(^3\!P_2) \Big] \bigg\} \,. \end{equation}
From the entries in Table\,IV of
ref.\cite{evgeni} we obtain the average values $C(\epsilon_1)= -3.95\,$fm$^4$ 
and $3C(^3\!P_1)-2 C(^3\!P_0)- C(^3\!P_2) =-20.5\,$fm$^4$, which will be used
for the numerical evaluation of eq.(40). Additional contributions to the
tensorial Landau parameters come from the diagrams in Fig.\,5 involving
$\Delta$-excitations and medium modifications. The diagram where the pion 
couples to delta-hole states leads to the contribution:
\begin{equation}  h_0(k_f) +  T\, h'_0(k_f) =  {g_A^4 k_f^3  \over 36
\pi^2 \Delta f_\pi^4 }(3-T)  \bigg[ \ln(1+4u^2)-{4u^2 \over 1+4u^2} \bigg] \,.
\end{equation} 
Somewhat more involved it the evaluation of the (second and third)
$2\pi$-exchange box diagrams in Fig.\,5 with Pauli blocking. We find the
following representation: 
\begin{eqnarray} h_0(k_f) +  T\, h'_0(k_f)& =&  {g_A^4 m_\pi^3 \over 96\pi^2 
\Delta f_\pi^4 }(3-T) \bigg\{ {5u^3 \over 3}+\int_0^u dx\, \bigg[-{2 L\over u} 
\Big((u^2-x^2)^2+u^2+x^2\Big )\nonumber  \\ && + {L^2 \over u} \Big(u(1+u^2
-x^2)^2+4x^2(x-u)(1-u x +x^2)\Big)\nonumber  \\ && + \int_{-1}^1 dy\,{ x^2
(1+u^2+x^2)\over 1+u^2+x^2 -2u x y} \, \ln{1+u^2+x^2 +2u x y \over
1+(u+x)^2}\bigg] \bigg\} \,,  \end{eqnarray} 
with the logarithmic auxiliary function:
\begin{equation} L ={1\over 4x}\ln{1+(u+x)^2 \over 1+(u-x)^2} \,.\end{equation}
The contribution of last pion-exchange diagram in Fig.\,5 with a density 
dependent vertex correction can be given in closed analytical form: 
\begin{eqnarray}   h_0(k_f) +  T\, h'_0(k_f) &=&  {g_A^4 m_\pi^3 \over 9\pi^2 
\Delta (4f_\pi)^4}(3-T) \bigg\{ -64 \ln(1+4u^2) \,\arctan 2u\nonumber \\  && 
-{9+33u^2-192u^4 +112 u^6 \over 3 u^3 } \ln(1+4u^2) \nonumber \\  &&+{2\over
u} (3+6u^2 -8u^4)+ {3+16u^2 +144 u^4 \over 8 u^5 } \ln^2(1+4u^2)  \bigg\} \,. 
\end{eqnarray}

\begin{figure}
\begin{center}
\includegraphics[scale=0.75]{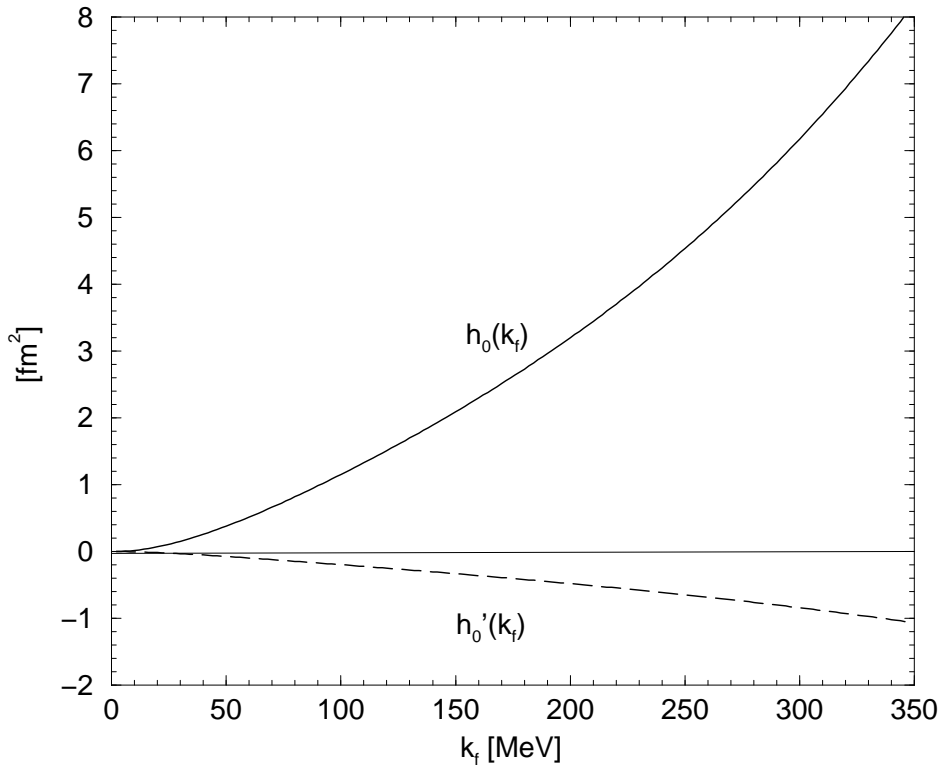}
\end{center}\vspace{-0.7cm}
\caption{The Landau parameters $h_0(k_f)$ and $h'_0(k_f)$ related to the
tensor interaction as a function of the Fermi momentum $k_f$. Contributions 
from two-pion exchange with excitation of virtual $\Delta(1232)$-isobars are 
included in addition.}    
\end{figure}

Fig.\,8 shows again the tensor Landau parameters $h_0(k_f)$ and $h'_0(k_f)$ 
as a function of the Fermi momentum $k_f$. We have now included the additional
contributions eqs.(39-44) from the chiral $\pi N\Delta$-dynamics and we have
kept the relativistic $k_f^2/M^2$-correction to the $1\pi$-exchange term 
eq.(33). The solid and dashed lines in Fig.\,8 for the isoscalar $h_0(k_f)$ 
and isovector $h_0'(k_f)$ tensor interaction strength are very similar to
the previous ones in Fig.\,7. At nuclear matter saturation density $k_{f0} =
263\,$MeV we find now $h_0(k_{f0}) = 4.93\,$fm$^2$ and $h_0'(k_{f0}) 
=-0.69\,$fm$^2$. In comparison to the (static) $1\pi$-exchange approximation
this amounts again to almost a doubling  of the tensor interaction strength in
the isoscalar channel, whereas the isovector channel is much less affected.  
The short-distance term eq.(40) contributes to the total result at the level 
of about $-11\%$. Its numerical value $(-0.56+ 0.08\, T)\,$fm$^2$ resembles 
the ''weak'' $\rho$-meson contribution of ref.\cite{brown} (see Table\,1
therein). We conclude therefore that the sizeable effects of iterated
$1\pi$-exchange on the tensor interaction at the Fermi surface are not altered
by higher order corrections from (irreducible) $2\pi$-exchange with virtual 
$\Delta$-isobar excitation and explicit short-distance contributions. 

It is well known that the expansion in Legendre-polynomials eq.(32) converges
slowly in case of the tensor interaction \cite{brown,dabr}. For that reason 
we have calculated also the $l=1$ tensor Fermi-liquid parameters $h_1(k_f)$
and $h'_1(k_f)$ in the present framework. The corresponding (numerical) 
results are shown as a function of the Fermi-momentum $k_f$ in Fig.\,9. The 
dashed lines correspond to the approximation to single and iterated
pion-exchange and the full lines include in addition the (higher order) 
contributions from $2\pi$-exchange with virtual $\Delta(1232)$-isobar 
excitation. In comparison the leading contribution from (static) 
$1\pi$-exchange:  
\begin{equation} h_1(k_f)^{1\pi} +  T\, h'_1(k_f)^{1\pi} ={g_A^2 \over
16f_\pi^2}(3-T) \Big[ (2+u^{-2})\ln(1+4u^2)-4\Big] \,, \end{equation}
one finds a pattern similar to the $l=0$ case. At nuclear matter saturation 
density $k_{f0}= 263\,$MeV one has now from static $1\pi$-exchange: 
$h_1(k_{f0})^{1\pi} +  T\, h'_1(k_{f0})^{1\pi}= (3.33 -1.11\,T)\,$fm$^2$. 
These numbers get enhanced to $h_1(k_{f0})+ T\, h'_1(k_{f0})= (5.08-1.83\,T)\,
$fm$^2$ by the iterated pion-exchange effects, whereas the further inclusion of
contributions from $2\pi$-exchange with virtual $\Delta(1232)$-excitation
leads only in minor changes: $h_1(k_{f0})+ T\, h'_1(k_{f0})= (4.37-1.85\,T)\,
$fm$^2$. One should also note that to the order in the small momentum
expansion we are working here, $h_1(k_f)$ and $h'_1(k_f)$ receive no
contribution from short-distance contact terms. We can therefore conclude that
the features observed for the $l=0$ tensor Fermi-liquid parameters are
approximately repeated at $l=1$ and presumably they continue to hold at higher 
$l$.

\begin{figure}
\begin{center}
\includegraphics[scale=0.7]{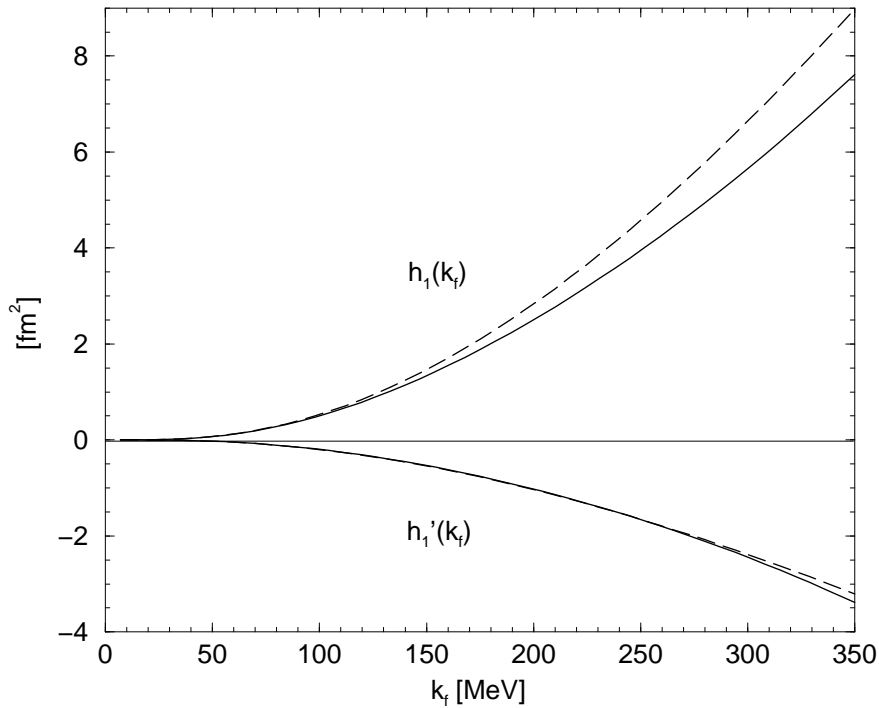}
\end{center}\vspace{-0.7cm}
\caption{The Landau parameters $h_1(k_f)$ and $h'_1(k_f)$ related to the
tensor interaction as a function of the Fermi momentum $k_f$. The dashed lines
show contributions from  single and iterated pion-exchange only. The full
lines include also contributions from two-pion exchange with virtual 
$\Delta(1232)$-isobar excitation.}    
\end{figure}

\section{Summary and conclusions}
We have explored in this work the role of the long-range $2\pi$-exchange for 
the in-medium interaction of quasi-nucleons at the Fermi surface $|\vec 
p_{1,2}|=k_f$. Our analytical calculation is rooted in a recent chiral 
approach to nuclear matter which can successfully explain binding and 
saturation of nuclear matter through two-pion exchange mechanisms
\cite{nucmat,deltamat}. The isotropic part of the quasi-nucleon interaction is
characterized by four density-dependent (dimensionful) Fermi-liquid parameters
$f_0(k_f),\,f_0'(k_f),\, g_0(k_f)$ and $g_0'(k_f)$. In the approximation to 
$1\pi$-exchange and iterated $1\pi$-exchange we find a spin-isospin
interaction strength of $g_0'(2m_\pi) = 1.14\, $fm$^2$, compatible with 
empirical values. The consistency relations to the nuclear matter
compressibility $K$ and the spin/isospin asymmetry energies are fulfilled in
our perturbative calculation. However, the non-linear Pauli principle sum
rules \cite{backman,schwenk} are most likely violated. 

In the next step we have included in the
quasi-particle interaction the contributions from $2\pi$-exchange with virtual
$\Delta(1232)$-isobar excitation. This extension to higher orders in the small
momentum expansion is necessary in order to obtain good single-particle 
properties and to guarantee spin-stability of nuclear matter \cite{spinstab}. 
Leaving out the short-distance terms (contributing proportional to $k_f^2$) the
spin-dependent Landau parameters $g_0(k_{f0})$ and $g'_0(k_{f0})$ come out far
too large. Estimates of these short-distance parameters from realistic 
NN-potentials go in the right direction, but sizeable enhancement factors are 
still needed to reproduce the empirical values of $g_0(k_{f0})\simeq 0.34\,
$fm$^2$  and $g_0'(k_{f0})\simeq 1.15$\,fm$^2$ \cite{schwenk}. Another 
shortcoming of the present calculation is that only leading one-loop 
pion-exchange contributions to the induced interaction \cite{babu,sjoberg,
brown,schwenk,wambach} could be included consistently. This is presumably the 
reason for our failure to describe the spin-dependent Fermi-liquid parameters,
which are not constrained by any bulk properties of nuclear matter. Actually,
one conclusion of ref.\cite{achim} has been that an accurate description of the
spin-response should include the induced interaction.

We have also considered the tensor part $3\,\vec \sigma_1 \cdot\vec q \,\vec 
\sigma_2\cdot \vec q -\vec \sigma_1 \cdot\vec \sigma_2\,{\vec q}^{\,2}$ of the
quasi-nucleon interaction at the Fermi surface. In comparison to the leading
$1\pi$-exchange  tensor interaction we have found from the iterated
$1\pi$-exchange corrections almost a doubling of the isoscalar tensor strength
$h_0(k_f)$, whereas the isovector tensor strength $h_0'(k_f)$ was much less
affected. This feature did not change qualitatively by the inclusion of the
chiral $\pi N\Delta$-dynamics. A similar pattern has been observed for the 
$l=1$ tensorial Fermi-liquid parameters $h_1(k_f)$ and $h_1'(k_f)$.  

\subsection{Acknowledgements}
I thank S. Fritsch for preparation of Fig.\,3 and W. Weise for informative
discussions.

\end{document}